\documentclass{article}
\usepackage{spconfa4,amsmath,graphicx}
\usepackage{booktabs}
\usepackage{url}
\usepackage{acronym}
\usepackage{caption}
\makeatletter
\def\section{\@startsection{section}{1}{\z@}%
  {1.2ex plus 0.4ex minus 0.2ex}
  {0.5ex plus 0.2ex}
  {\normalfont\normalsize\bfseries\centering}}

\def\subsection{\@startsection{subsection}{2}{\z@}%
  {2.0ex}
  {1ex}
  {\normalfont\normalsize\bfseries}}
\makeatother

\title{DNN-Based Frequency-Dependent Estimation of Speech, Music, and Noise Power in Acoustic Mixtures for Hearing-Aid Scene Analysis}
\name{Mats Lang$^{1,2}$, Thomas Haubner$^{2}$, Nina Kiessling$^{2}$, Christoph Hoog Antink$^{1}$, Henning Puder$^{1,2}$}
\address{$^{1}$Technische Universität Darmstadt, Darmstadt, Germany \\ $^{2}$WSA, Erlangen, Germany}
\begin{document}
%
\acrodef{DNN}{Deep Neural Network}
\acrodef{RMS}{Root Mean Square}
\acrodef{CRNN}{Convolutional Recurrent Neural Network}
\acrodef{GRU}{Gated Recurrent Unit}
\acrodef{VAD}{Voice Activity Detection}
\acrodef{SNR}{Signal-to-Noise Ratio}
\acrodef{STFT}{short-time Fourier transform}
\acrodef{ReLU}{Rectified Linear Unit}
\acrodef{MSE}{Mean Squared Error}
\acrodef{MAE}{Mean Absolute Error}
\acrodefplural{RIR}[RIRs]{Room Impulse Responses}
\acrodef{RIR}[RIR]{Room Impulse Response}
\acrodef{HRIR}[HRIR]{Head-Related Impulse Response}
\acrodef{ROC}{Receiver Operating Characteristic}
\acrodef{LE}{Level Error}
\acrodef{PCC}{Pearson Correlation Coefficient}
\acrodef{PSD}{Power Spectral Density}
\acrodef{IIR}{Infinite Impulse Response}
\acrodef{GPU}{Graphic Processing Unit}
\acrodef{SOTA}{state-of-the-art}

\maketitle
%

\begin{abstract}
Acoustic scene analysis is essential for adapting hearing-aid signal processing algorithms to the current listening environment. However, \ac{SOTA} systems typically rely on multiple independent estimators for tasks such as scene classification, \ac{VAD}, or \acl{SNR} estimation, which increases computational complexity and fails to exploit dependencies between related tasks. To address this problem, we propose a unified and interpretable acoustic scene representation by decomposing the observed mixture spectrum into speech, music, and noise power components. This is motivated by the typical listening targets of hearing-aid users. In particular, we estimate time- and frequency-dependent power proportions by a causal low-complexity \acl{DNN}, from which multiple downstream acoustic scene analysis measures can in principle be derived by simple post-processing. In this work, we validate the proposed representation using \ac{VAD} as a representative downstream task and show performance comparable to a \ac{SOTA} estimator while providing a substantially richer scene description.
\end{abstract}

\acresetall

\section{Introduction}
\label{sec:intro}
It is well known that hearing-aid users benefit from algorithms that adapt to the current acoustic scene, which requires reliable acoustic scene analysis \cite{launerHearingAidSignal2016b}.
In current hearing aids, \ac{SOTA} acoustic scene analysis is typically implemented by multiple independent estimators, for example for acoustic scene classification~\cite{gil-pitaComputationallyEfficientSound2015}, \ac{VAD}~\cite{liuComputationEfficientVoice2021}, \ac{SNR} estimation~\cite{mayAssessmentBroadbandSNR2017b}, and source localization or directional analysis~\cite{zohourianBinauralSpeakerLocalization2018}. Their outputs are used to control functions such as compression, noise reduction, and automatic program selection~\cite{searchfieldPerformanceAutomaticAcousticbased2018}. This modular design has two main drawbacks: First, it is computationally inefficient, which becomes increasingly critical as computationally-demanding deep-learning methods are introduced into modern hearing aids~\cite{wangDeepLearningReinvents2017}. Second, independent estimators do not exploit dependencies between related tasks, although they often rely on overlapping information extracted from the same acoustic mixture, such as in \ac{VAD} and \ac{SNR} estimation.

To address these limitations, we propose a unified and interpretable acoustic scene representation from which multiple relevant analysis estimates can be obtained by simple post-processing. Specifically, the mixture spectrum is decomposed into speech, music, and noise power components. These three classes are of particular relevance in hearing-aid applications, since speech and music are typically desired signals, whereas other undesired components can be grouped as noise. The proposed decomposition is realized by a low-complexity causal \ac{DNN} that estimates class-wise, time- and frequency-dependent relative power proportions, which are subsequently combined with the mixture power to obtain class-specific time-varying power spectrum estimates. This relative parametrization is, in principle, independent of the absolute input level while still enabling power estimation. In contrast to hard scene labels, as employed in acoustic scene classification, the proposed representation preserves simultaneous and frequency-dependent class activity while remaining substantially simpler than full source separation, as commonly considered in speech enhancement~\cite{nasimAudioSourceSeparation2025}.

The proposed representation can in principle support multiple downstream tasks, including \ac{VAD}, \ac{SNR} estimation, and acoustic scene classification, within a common framework. This may reduce overall computational complexity while enabling the model to exploit dependencies between related tasks. In the present work, we investigate the estimation accuracy, generalization, interpretability, and complexity of the proposed approach, and assess its downstream usefulness for \ac{VAD} as a representative task.


\section{Proposed Method}
\label{sec:proposed_method}
This section presents the proposed acoustic scene analysis method, including the target definition, the DNN-based estimator, and the training loss.

\subsection{Regression target}
\label{sec:signal_model_target}
The target representation is defined on a time-frequency decomposition, where $k$ is the frequency-band index and $\ell$ is the time-frame index.\\
The observed mixture signal is modeled as
\begin{align}
    X(k,\ell) = S(k,\ell) + M(k,\ell) + N(k,\ell),
\end{align}
where $S(k,\ell)$, $M(k,\ell)$, and $N(k,\ell)$ denote the composite speech, music, and noise components, respectively, each comprising all active sources of the corresponding class present in the mixture.\\ Let $C_{\mathrm{c}}(k,\ell) \in \{S(k,\ell), M(k,\ell), N(k,\ell)\}$ denote a generic class-specific component, and define the corresponding power as
\begin{align}
    P_{\mathrm{c}}(k,\ell) = |C_{\mathrm{c}}(k,\ell)|^2, \qquad \mathrm{c} \in \{\mathrm{s},\mathrm{m},\mathrm{n}\}.
\end{align}
Analogously, the mixture power is given by $P_{\mathrm{x}}(k,\ell) = |X(k,\ell)|^2$. Under the assumption of mutually uncorrelated components, $P_{\mathrm{x}}(k,\ell)$ equals the sum of class-specific powers only in expectation. We therefore define
\begin{align}
    P_{\Sigma}(k,\ell) = \sum_{\mathrm{c} \in \{\mathrm{s},\mathrm{m},\mathrm{n}\}} P_{\mathrm{c}}(k,\ell)
\end{align}
and the relative power proportion of class $\mathrm{c}$ as
\begin{align}
    r_{\mathrm{c}}(k,\ell) = \frac{P_{\mathrm{c}}(k,\ell)}{P_{\Sigma}(k,\ell)}.
    \label{eq:relative_power}
\end{align}
These quantities sum to one in each time-frequency bin and are independent of the absolute signal level. Since during inference $P_{\Sigma}(k,\ell)$ is unavailable, class-specific powers are approximated from the estimated proportions $\hat{r}_{\mathrm{c}}(k,\ell)$ and the observable mixture power via
\begin{align}
    \hat{P}_{\mathrm{c}}(k,\ell) = \hat{r}_{\mathrm{c}}(k,\ell)\, P_{\mathrm{x}}(k,\ell).
    \label{eq:power_reconstruction}
\end{align}

\subsection{DNN-based estimator and training objective}
\label{sec:dnn_estimator}
\begin{figure}
    \hspace*{-0.08\linewidth}
    \includegraphics[width=1.3\linewidth]{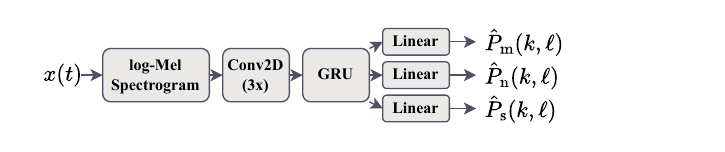}
    \caption{Architecture of the proposed causal \ac{CRNN} for estimating time-varying speech, music, and noise power spectra. $x(t)$ denotes the time-domain mixture.}
    \label{fig:crnn_architecture}
\end{figure}

In the implemented system, the proposed estimator operates on causal \ac{RMS}-normalized log-Mel spectrogram features obtained by applying a Mel filterbank to the mixture \ac{STFT} power spectrum followed by logarithmic compression \cite{1163420}. The input feature in Mel band $b$ at time frame $\ell$ is defined as
\begin{align}
    f_b(\ell) = \log \left( \sum_{q=1}^{Q} H_{b,q} P_\mathrm{x}(q,\ell) + \epsilon \right),
\end{align}
where $b \in \{1,\dots,B\}$ is the Mel-band index, $B$ is the number of Mel bands, $q$ is the \ac{STFT} frequency-bin index, $Q$ is the number of \ac{STFT} frequency bins, $H_{b,q}$ is the Mel filterbank coefficient, and $\epsilon > 0$ is a small constant for numerical stability. Prior to logarithmic compression, the Mel-band energies are normalized frame-wise by a causal temporally smoothed \ac{RMS} estimate.
The estimator is a lightweight causal \ac{CRNN} that combines convolutional layers for local spectro-temporal feature extraction with a recurrent layer for temporal modeling~\cite{adavanneSoundEventLocalization2019}. Its architecture is shown in Fig.~\ref{fig:crnn_architecture}. The input log-Mel spectrogram is processed by three 2-D convolutional layers, each followed by batch normalization, \ac{ReLU} activation, and dropout. The resulting feature maps are flattened along the channel and frequency dimensions and passed to a causal unidirectional \ac{GRU}, whose output is fed to three separate fully connected layers corresponding to speech, music, and noise. For each time-frequency bin, the resulting three logits are normalized by a softmax over the class dimension, yielding the estimated relative class proportions as defined in~\eqref{eq:relative_power}. The class-specific power estimates are then obtained according to~\eqref{eq:power_reconstruction}. Causality is ensured by left-only padding in the time dimension.

Training is based on the empirically motivated point-wise penalty in~\eqref{eq:pointwise_penalty}, defined on the estimated and target class-specific powers:
\begin{align}
    \epsilon_{\mathrm{c}}(k,\ell)
    &=
    \log\left(
        \frac{1}{\bar{P}_{\Sigma}}\,
        \left| \hat{P}_{\mathrm{c}}(k,\ell) - P_{\mathrm{c}}(k,\ell) \right|
        + 1
    \right),
    \label{eq:pointwise_penalty}
\end{align}
where the normalization term $\bar{P}_\Sigma$ is given by
\begin{align}
    \bar{P}_{\Sigma}
    =
    \frac{1}{KL}
    \sum_{\ell=1}^{L}
    \sum_{k=1}^{K}
    P_{\Sigma}(k,\ell),
    \label{eq:mean_power}
\end{align}
i.e., the mean summed source power of the corresponding training example, with $K$ frequency bands and $L$ time frames. The overall loss is obtained by averaging \eqref{eq:pointwise_penalty} over all classes, frequency bands, and time frames. In~\eqref{eq:pointwise_penalty}, normalization by~\eqref{eq:mean_power} reduces sensitivity to the overall signal level while still emphasizing dominant regions in linear power. The offset in~\eqref{eq:pointwise_penalty} ensures a non-negative logarithm argument and yields zero penalty for perfect estimation.

\section{Experiments}
\label{sec:experimental_setup}
In the following, we describe the experimental evaluation of the proposed algorithm.

\subsection{Data generation}
\label{sec:data}
The training, validation and testing data was generated from a variety of publicly-available speech, music, and noise datasets. Speech signals were taken from LibriSpeech~\cite{panayotovLibrispeechASRCorpus2015} and VCTK~\cite{yamagishiCSTRVCTKCorpus2019}, music from MUSAN~\cite{snyderMUSANMusicSpeech2015} and FMA~\cite{defferrardFMADatasetMusic2017}, and noise from MUSAN~\cite{snyderMUSANMusicSpeech2015}, DNS~\cite{dubeyICASSP2022Deep2022}, DNC~\cite{rafaelDNCDatasetNoise2021}, ESC-50~\cite{piczakESCDatasetEnvironmental2015}, UrbanSound8K~\cite{salamonDatasetTaxonomyUrban2014}, and DEMAND~\cite{thiemannDiverseEnvironmentsMultichannel2013}. Datasets without predefined splits were partitioned into 80\%, 10\%, and 10\% for training, validation, and test.

 Mixtures were generated with equal ratios of jointly active N, M, S, N+M, N+S, M+S, and N+M+S, where N, M, and S denote noise, music, and speech. For each active class, a signal excerpt was randomly sampled from the respective publicly available dataset, resampled to \(16\,\mathrm{kHz}\), normalized to unit variance, and scaled by a randomly sampled level offset in the range \([-10,10]\,\mathrm{dB}\). In addition, low-level background noise from DEMAND was added with a power randomly sampled from \([-40,-20]\,\mathrm{dB}\). To increase variability, augmentations were applied to the individual source signals prior to mixing. Reverberation was applied with probability $2/3$ by convolving each source signal with a randomly sampled \ac{RIR} from the MIT database \cite{traerStatisticsNaturalReverberation2016}. Music signals were further modified by three-band equalization, speech signals by a band-pass filter, and all source types optionally by a linear gain drift of up to \(\pm 3\,\mathrm{dB}\) to simulate slow level variations commonly encountered in real-world recordings. The development dataset consisted of 50,000 training, 6,250 validation, and 6,250 test samples of length $6\,\mathrm{s}$.

For the unseen-data evaluation, noise and music signals were taken from the HEAR-DS hearing-aid recordings~\cite{huwelHearingAidResearch2020}, while speech signals were taken from TIMIT~\cite{timit_dataset}. The same class combinations as in the training data were used. Following the HEAR-DS evaluation setup, the TIMIT utterances were convolved with \acp{HRIR}~\cite{thiemannMultipleModelHighresolution2019} to obtain binaural speech signals compatible with the hearing-aid recording scenario. These speech signals were then mixed with the HEAR-DS noise and music signals using relative level differences randomly sampled from \([-10,10]\,\mathrm{dB}\).

\subsection{Algorithmic settings}
\label{sec:implementation_details}

The \ac{STFT} was computed using a window length of 400 samples and a hop size of 200 samples, corresponding to a 25\,ms window and a 12.5\,ms frame shift at 16\,kHz. From the mixture signal, $B=64$ log-Mel bands were computed as feature and causally \ac{RMS}-normalized with a first-order \ac{IIR} filter with time constant 1 s. Regression targets were generated on \(K=16\) Mel bands and temporally smoothed by a first-order \ac{IIR} filtering with a time constant of \(0.1\,\mathrm{s}\) to reduce strong short-term power fluctuations, which can otherwise hinder stable model optimization. The convolutional layers used 8, 16, and 24 output channels with $3\times3$ kernels; the first two layers had unit stride, and the third used stride $(2,1)$, where the downsampling is applied along the frequency dimension only. Dropout was set to 0.1, and the unidirectional \ac{GRU} used 64 hidden units. The three output layers produced $K=16$ estimates per class, corresponding to the 16 Mel bands of the target representation. Training samples had a duration of 6\,s, corresponding to $L=480$ time frames. Training was performed in PyTorch using Adam with learning rate $5 \times 10^{-3}$, weight decay $10^{-5}$, batch size 64, and gradient clipping with norm 0.5.

\subsection{Performance metrics}
\label{sec:evaluation_metrics_setup}

Performance was evaluated by comparing the estimated class-specific power spectra to the corresponding target powers for each class as defined in \eqref{eq:power_reconstruction}. The logarithmic \ac{LE} per class $c$ was defined as
\begin{align}
    E_{\mathrm{LE},\mathrm{c}}
    &= \frac{1}{KL}
    \sum_{\ell=1}^{L}
    \sum_{k=1}^{K}
    \left|
    10 \log_{10}\!\left(
    \frac{\hat{P}_{\mathrm{c}}(k,\ell) + \delta(k,\ell)}{P_{\mathrm{c}}(k,\ell) + \delta(k,\ell)}
    \right)
    \right|,
    \label{eq:level_error}
\end{align}
where
\begin{align}
    \delta(k,\ell) = 0.03\,P_{\mathrm{x}}(k,\ell)
\end{align}
is a heuristic power floor chosen empirically to stabilize the logarithmic error in low-power bins. It corresponds to approximately $15.2\,\mathrm{dB}$ below the mixture power in each time-frequency bin.

Linear agreement between estimate and target was assessed using the \ac{PCC} \cite{pearson1895}. For each class, the PCC was computed separately for each frequency band over time and then averaged across frequency bands. The \ac{LE} measures absolute deviations on a decibel scale, whereas the \ac{PCC} reflects the linear association between estimate and target.

\subsection{Results}
Except for the qualitative example, all results reported below were obtained on 1,000 unseen samples of 10 s duration.

\begin{figure}
    \centering
    \includegraphics[width=\linewidth]{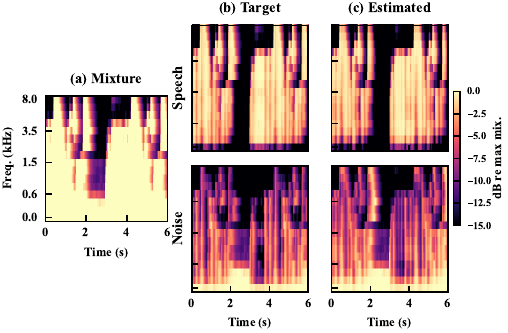}
    \caption{Qualitative example from the unseen evaluation dataset for speech mixed with \textit{InVehicle} noise. The left column shows the mixture, while the first and second rows show the target and estimated speech and noise levels, respectively.}
    \label{fig:online_example_noise_levels}
\end{figure}

A qualitative example was first used to assess whether the proposed method captures frequency-dependent source levels on unseen data. Fig.~\ref{fig:online_example_noise_levels} shows speech mixed with \textit{InVehicle} noise from HEAR-DS at 0\,dB \ac{SNR}. The left column shows the mixture, and the first and second rows compare the target and estimated speech and noise levels, respectively. The \textit{InVehicle} noise is concentrated mainly in the low-frequency bands and overlaps little with speech-dominant regions. The estimates follow this structure closely, indicating that the model captures local spectro-temporal differences between speech and interference. In contrast to hard labels, the proposed representation preserves such frequency-dependent dominance patterns.

Next, overall regression performance was evaluated on the development and unseen datasets using \ac{LE} and \ac{PCC}, aggregated over all classes, frequency bands, and time frames. The proposed model achieved an overall \ac{LE} of 1.40\,dB and a \ac{PCC} of 0.891 on the development set, and 1.71\,dB and 0.875 on the unseen set. These results indicate good agreement between estimated and target power spectra. A slight degradation in performance was observed on the unseen dataset.

\begin{figure}
    \centering
    \includegraphics[width=\linewidth]{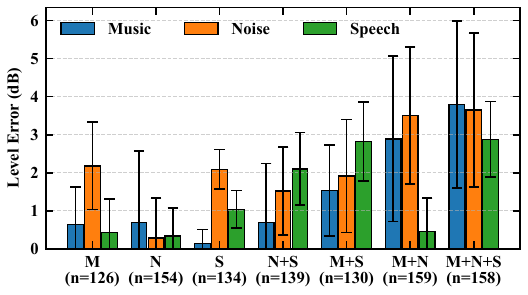}
    \caption{Mean and standard deviation of \ac{LE} (dB) (cf.~\eqref{eq:level_error}) as a function of source-type combination. S, M, and N denote speech, music, and noise, respectively.}
    \label{fig:regression_dataset_combination_level_error}
\end{figure}

To analyze performance across acoustic conditions, \ac{LE} was further evaluated by source-type combination on the unseen dataset. As described in Section~\ref{sec:data}, low-level background noise was added to all samples, including the nominally single-source conditions. Fig.~\ref{fig:regression_dataset_combination_level_error} therefore shows that, in single-source conditions, the remaining errors mainly arise from assigning energy to absent classes. For speech-only and music-only samples, the largest error occurs for the noise class, indicating that low-energy or ambiguous regions are occasionally attributed to noise. Noise-only samples yield the lowest overall error, suggesting that this condition is estimated most reliably. For mixtures, the error increases with the number of active source types. Among two-source mixtures, noise-plus-speech and music-plus-speech yield lower errors than music-plus-noise, while the three-source case gives the largest error. This suggests that speech is easier to distinguish, likely due to its more structured spectral characteristics, whereas music and noise overlap more strongly.

\begin{figure}
    \centering
    \includegraphics[width=0.85\linewidth]{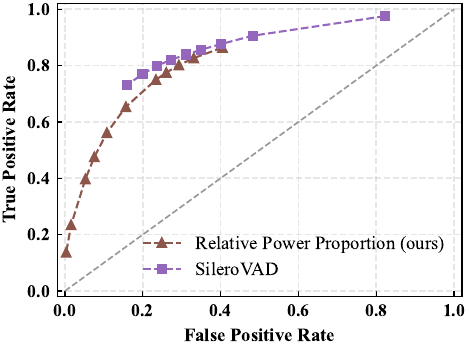}
    \caption{\ac{ROC} curve for \ac{VAD} results of Silero\ac{VAD} \cite{SileroVAD2024} and the proposed relative power proportion-based method.}
    \label{fig:vad_algorithm_roc_curve}
\end{figure}

The representation was also evaluated in a downstream \ac{VAD} task using the estimated relative speech proportion \(\hat{r}_\mathrm{s}(k,\ell)\) from \eqref{eq:relative_power} as a level-independent speech-presence cue. Frame-level targets were derived from the clean-speech energy, \(\hat{r}_\mathrm{s}(k,\ell)\) was averaged across frequency, and performance was measured using \ac{ROC} curves and balanced accuracy, i.e., the average of sensitivity and specificity. The proposed method was compared with Silero\ac{VAD} \cite{SileroVAD2024}, a \ac{DNN}-based \ac{SOTA} method. Silero\ac{VAD} thresholds from 0.1 to 0.9 and power-proportion thresholds from 0.01 to 0.9 were evaluated. As shown in Fig.~\ref{fig:vad_algorithm_roc_curve}, the proposed method performs comparably to Silero\ac{VAD} in the upper-left region of the \ac{ROC} space. At the selected operating points, the proposed method achieves a balanced accuracy of 0.845, compared with 0.848 for Silero\ac{VAD}. Thus, near-\ac{SOTA} \ac{VAD} performance can be obtained by simple thresholding of the estimated relative speech power, while retaining frequency-dependent information that is unavailable in binary activity labels.

Finally, practical feasibility was assessed in terms of model size and complexity. The proposed \ac{CRNN} comprises 181.9k parameters, corresponding to approximately 727.6\,kB for 32-bit weights, and requires 30.9\,M floating-point operations per second, supporting its suitability for low-complexity hearing-aid applications.

\section{Conclusion and Outlook}
\label{sec:conclusion_outlook}
This paper presented a low-complexity causal \ac{CRNN} for decomposing the observed mixture spectrum into speech, music, and noise power components. Results on development and unseen evaluation data showed promising estimation accuracy, while also indicating increased errors for more complex acoustic scenarios involving multiple source types. A qualitative example illustrated the benefit of frequency-dependent estimation beyond hard scene labels, and a downstream \ac{VAD} experiment demonstrated competitive performance. Overall, the proposed representation appears to be a promising low-complexity basis for downstream hearing-aid analysis tasks. Future work will compare the proposed representation in additional downstream tasks, investigate its combination with beamforming for localization-related analysis, and further reduce computational complexity.



\begingroup
\small
\bibliographystyle{IEEEbib}
\bibliography{refs}
\endgroup

\end{document}